\documentclass{article}

\usepackage{PRIMEarxiv}

\usepackage[utf8]{inputenc} 
\usepackage[T1]{fontenc}    
\usepackage{hyperref}       
\usepackage{url}            
\usepackage{booktabs}       
\usepackage{amsfonts}       
\usepackage{nicefrac}       
\usepackage{microtype}      
\usepackage{lipsum}
\usepackage{fancyhdr}       
\usepackage{graphicx}       
\graphicspath{{media/}}     
\usepackage{float}
\usepackage{bm}
\usepackage{amsmath}
\usepackage{caption}
\usepackage{subcaption}
\usepackage{amssymb}
\usepackage{placeins}
\title{Open Access Improves the Dissemination of Science: Insights from Wikipedia}

\author{
  Puyu Yang\\
  Institute for Logic, Language and Computation (ILLC) \\
  The University of Amsterdam \\
  The Netherlands\\
  \texttt{p.yang2@uva.nl} \\
   \And
  Ahad Shoaib \\
  School of Computer and Communication Sciences \\
  École Polytechnique Fédérale de Lausanne (EPFL) \\
  Switzerland\\
  University of Waterloo\\
    Canada\\
   \AND
   Robert West \\
   School of Computer and Communication Sciences \\
  École Polytechnique Fédérale de Lausanne (EPFL) \\
  Switzerland\\
   \And
   Giovanni Colavizza \\
   Department of Classical Philology and Italian Studies \\
  University of Bologna \\
  Italy\\
}

\begin{document}
\maketitle

\begin{abstract}
Wikipedia is a well-known platform for disseminating knowledge, and scientific sources, such as journal articles, play a critical role in supporting its mission. The open access movement aims to make scientific knowledge openly available, and we might intuitively expect open access to help further Wikipedia's mission. However, the extent of this relationship remains largely unknown.
To fill this gap, we analyze a large dataset of citations from the English Wikipedia and model the role of open access in Wikipedia's citation patterns. 
We find that both the accessibility (open access status) and academic impact (citation count) significantly increase the probability of an article being cited on Wikipedia.
Specifically, open-access articles are extensively and increasingly more cited in Wikipedia, as they show an approximately 64.7\% higher likelihood of being cited in Wikipedia when compared to closed-access articles, after controlling for confounding factors. This open-access citation effect is particularly strong for articles with high citation counts and published in recent years.
Our findings highlight the pivotal role of open access in facilitating the dissemination of scientific knowledge, thereby increasing the likelihood of open-access articles reaching a more diverse audience through platforms such as Wikipedia. Simultaneously, open-access articles contribute to the reliability of Wikipedia as a source by affording editors timely access to novel results.
\end{abstract}

\keywords{Wikipedia \and Open Access \and Open Science \and Science Communication}

\section{Introduction}
Open access (OA) publishing has emerged as a popular alternative to traditional subscription-based models, with the goal of making research more widely accessible to the public. This movement has gained momentum over the years, with many scholars recognizing the benefits of open access in promoting the dissemination of scientific knowledge and funding bodies adopting OA mandates~\cite{piwowar_state_2018, holmberg_articles_2020}.

Citations play a crucial role in supporting Wikipedia's mission to offer verifiable and reliable information\footnote{\url{https://en.wikipedia.org/wiki/Wikipedia:Core_content_policies}.}. Among various sources, academic and peer-reviewed publications are widely regarded as the most reliable\footnote{\url{https://en.wikipedia.org/wiki/Wikipedia:Verifiability\#Reliable_sources}.}. The open-access movement presents a significant opportunity for Wikipedia to access a vast repository of reliable and verifiable scientific knowledge. As a dynamic platform for sharing and disseminating knowledge across the globe, Wikipedia is relied upon by millions of users every day to satisfy a wide range of information needs~\cite{singer_why_2017}. It has become a critical source of information for both the general public and academic researchers, and its impact is extending beyond the realm of general knowledge and into the academic sphere~\cite{park_visibility_2011, kousha_are_2015, tohidinasab_why_2013}.
Wikipedia's extensive use of citations makes it possible to analyze its reliance on academic publications, which is a central aspect of our investigation. 
Previous research utilized the Scopus database and an English Wikipedia database dump extracted from 2014, culminating in the identification of 32,361 unique articles for analysis. They found that articles from open-access journals exhibit 47\% higher odds of being cited in Wikipedia compared to those from paywalled journals~\cite{teplitskiy_amplifying_2017}. Notably, their adoption of journals as the unit of analysis, rather than individual articles, has the possible drawback of wrongly estimating the influence of open access on scientific knowledge dissemination through Wikipedia. This limitation also arises from overlooking articles accessible via green or hybrid routes~\cite{elsabry2017needs}. Moreover, their manual matching approach imposed constraints on the scale of their research. Therefore, an exploration conducted at the granularity of individual articles not only promises a more nuanced understanding of the relationship between open access and the dissemination of scientific knowledge through Wikipedia but also unveils the role of citation count in this process.

In light of this, our study seeks to fill this gap by examining how open-access publications affect Wikipedia at the article-level granularity. Specifically, we aim to answer the following research questions:
\begin{enumerate}
\item RQ1: To what extent does Wikipedia rely on open-access publications? How has this been changing over time?
\item RQ2: To what extent does the open-access status of an article influence its likelihood of being cited in Wikipedia?
\end{enumerate}

To address these questions, we will use descriptive statistics and regression analysis based on the \textit{Wikipedia Citations} dataset~\cite{halfaker_2018}. To identify the information in article-level granularity such as the open access status of publications, we will use the OpenAlex and Scimago data. Our research contributes to the understanding of the role of open access in the dissemination of scientific knowledge and the impact of Wikipedia in this process, as well as informing policy and practice in the realm of open scholarly communication.

The remainder of the paper is structured as follows. Section 2 provides an overview of existing research in the field. Section 3 describes our dataset and methodology. Section 4 presents descriptive statistics of open-access publications in Wikipedia (RQ1), and then uses regression analysis to model the influence of open-access status on the likelihood of a paper being cited in Wikipedia (RQ2). Finally, Section 5 and 6 offer a discussion and conclusion of our findings.

\section{Previous Work}

\subsection{Open Access in Science}
The key idea behind open access (OA) is to provide unrestricted and free access to scientific outcomes, thus enhancing their visibility and reach regardless of financial or geographical constraints~\cite{tennant_academic_2016,redalyc2003berlin}. The increasing popularity of OA in academic publications has generated extensive discussions among scholars in recent years. Empirical studies have shown that OA has had a significantly positive impact on the accessibility of scientific journal articles~\cite{bjork_open_2010}. A comprehensive analysis of OA publications shows that at least 27.9\% of the total 19 million scientific articles are OA~\cite{piwowar_state_2018}. In addition, studies report that around 55\% of articles indexed by the Web of Science from 2009 to 2014 are OA, and more than 50\% of scientific papers published since 2007 can be accessed freely~\cite{martin-martin_evidence_2018,archambault_proportion_2014}. Among the various OA policies, Bronze OA is the most common type~\cite{piwowar_state_2018}. Although the distribution of OA varies across different fields, General Science, Technology, and Biomedical research have relatively higher OA rates, while Engineering and Arts\&Humanities have lower rates~\cite{archambault_proportion_2014,martin-martin_evidence_2018}.

An ``open access citation advantage" (OACA) has also been a topic of ongoing debate. Some researchers have observed that a citation advantage linked to open access exists, although the effect magnitude varies based on the dataset and methods used. For example, OA articles have been found to receive 18\% more citations than average based on Web of Science, while Scopus reports an even higher, positive 40\% effect~\cite{piwowar_state_2018,archambault_proportion_2014}. Kristin found that in four disciplines—philosophy, political science, electrical and electronic engineering, and mathematics, OA articles exhibit a greater research impact~\cite{antelman2004open}. Distinct advantages are found for green OA articles hosted in institutional repositories, receiving 106\% more citations than gold OA or non-OA articles, and OA articles obtained up to 36\% more diverse, interdisciplinary citations than non-OA articles~\cite{young2020green}. Despite these findings, a recent systematic review of OACA suggests that the debate continues, revealing diverse outcomes in different studies~\cite{langham2021open}. Out of 134 included studies, 47.8\% confirm the existence of OACA, 27.6\% deny it, 23.9\% find OACA only in subsets, and 0.8\% are inconclusive, with a notable association between the focus on multiple disciplines and the identification of OACA in subsets~\cite{langham2021open}. Therefore, the effects of OA on citation patterns remain a topic of interest and active investigation.

\subsection{Science and Wikipedia}
With the rapid development of the internet, traditional peer review processes require adaptation to align with the rapid knowledge creation in the 21st century~\cite{black_wikipedia_2008}. As one of the largest encyclopedias worldwide, Wikipedia aims to effectively and globally distribute information based on scientific findings\footnote{\url{https://wikimediafoundation.org/about/mission/}}, thereby making it a valuable altmetric source~\cite{sugimoto_scholarly_2017, mesgari2015sum}. Evans and Krauthammer observed higher citation counts for articles linked in Wikipedia, suggesting its potential for impact assessment~\cite{evans2011exploring}. Altmetric.com integrated Wikipedia mentions into its tracking in 2015\footnote{{\url{https://www.altmetric.com/blog/new-source-alert-wikipedia/}}}, but doubts arose about Wikipedia's reliability for impact assessment. Lin and Fenner found only 4\% of PLOS articles cited in Wikipedia~\cite{lin2014analysis}, and Kousha and Thelwall concluded that Wikipedia citations are insufficient for impact assessment in most fields~\cite{kousha_are_2015}. 

Previous research has indicated that the topical coverage of Wikipedia bears some similarity to science. With 13.44\% of its citations being from open access journals~\cite{arroyo2020science}, And 31.2\% of Wikipedia citations were associated with an Open Access (OA) source, this percentage exhibited an upward trend by year~\cite{pooladian2017methodological}. Additionally, STEM fields, especially biology and medicine, comprise the most prominently featured scientific topics in Wikipedia~\cite{yang2021map}. Specific fields such as Medicine and Psychology have a comparatively high number of citations to research papers on Wikipedia and are sometimes employed as a gateway to further academic research~\cite{maggio_wikipedia_2017,schweitzer_wikipedia_2008}. Furthermore, journal articles cited from Wikipedia tend to be published in high-impact journals (e.g., by impact factor) and in open access more frequently than the average article~\cite{nielsen_scientific_2007,teplitskiy_amplifying_2017}.

Science contributes a lot to Wikipedia, yet the influence goes both ways. Prior studies have established that Wikipedia can enhance the citation impact of an article it cites~\cite{thompson_science_2018}. Additionally, Wikipedia has demonstrated its ability to rapidly and reliably incorporate novel scientific findings in response to ongoing public events or crises~\cite{colavizza_covid-19_2020}.

\subsection{Citation Analyses of Wikipedia}
The open release of citation datasets from Wikipedia has led to a surge of studies examining citation analysis on Wikipedia~\cite{halfaker_2018,zagovora2020updated}. Among the articles on Wikipedia, 6.7\% cite at least one journal article with an associated DOI~\cite{halfaker_2018}, with the majority of cited journal articles being published in the past two decades~\cite{yang2021map}. \cite{benjakob_citation_2022} have conducted a study on the quality of citations in Wikipedia during COVID-19 and have found that Wikipedia mostly cites reliable sources and prefers open-access articles. Some researchers have focused on user behavior regarding reference usage on Wikipedia. \cite{piccardi_quantifying_2020} have found that engagement with citations on Wikipedia is generally low, but references are more frequently looked up when the information is not included.

Despite the increasing number of citation studies on Wikipedia, the relationship between open access and Wikipedia requires further exploration. Previous research has examined the effect of OA on Wikipedia, and found that articles with OA were 47\% more likely to be cited than Closed Access articles when controlling for journal and research fields~\cite{teplitskiy_amplifying_2017}. However, their emphasis on analyzing journals rather than individual articles results in an underestimation of open access's impact on disseminating scientific knowledge through Wikipedia. This limitation is due to the oversight of articles accessible through green or hybrid routes~\cite{elsabry2017needs}. Furthermore, their manual matching approach limited the scope of the research. Consequently, this study aims to improve upon previous findings by employing a more rigorous and comprehensive methodology, examining individual articles, and accounting for additional confounding factors better to comprehend the relationship between open access and Wikipedia.



\section{Data and methods}
The data collection process followed a specific workflow as outlined below. Firstly, we obtain all citations from English Wikipedia to any source using the open dataset known as \textit{Wikipedia Citations}~\cite{kokash_2024_10782978}. Additionally, to identify journal articles, we relied on the classification and DOI provided by \textit{Wikipedia Citations}. Secondly, to enrich the journal articles with article-level data, such as citation counts, open-access status and open-access policy, we employed the OpenAlex API to retrieve the relevant information through DOIs for each journal article. Finally, we utilized data from Scimago to acquire pertinent information for each journal. The subsequent sections provide a detailed description of the main datasets used in the study.

\subsection{Wikipedia Citations}
The primary dataset utilized in this research is \textit{Wikipedia Citations}, a comprehensive dataset of over 45 million citations extracted from the English Wikipedia February 2024 dump~\cite{kokash_2024_10782978}, which is an updated version based on the 2020 version~\cite{halfaker_2018}. Out of these, around 2.2 million citations are classified as journal articles, with 2,197,461 of them containing a digital object identifier (DOI).

\subsection{OpenAlex and Scimago}
To examine the impact of open access (OA) articles, we utilized OpenAlex, a free and open platform that provides data on academic papers and researchers~\cite{priem_openalex_2022}. OpenAlex draws data from Microsoft Academic Service (MAG) and Crossref, among other sources, and contains more than 240 million academic works that can be utilized in the fields of bibliometrics, science and technology studies, and science of science policy~\cite{bredahl_chapter_2022, hao_thirty-two_2022}. To obtain the necessary data for journal articles in \textit{Wikipedia Citations}, we utilized the OpenAlex API\footnote{\url{https://docs.openalex.org/how-to-use-the-api/get-single-entities}.} to obtain relevant article details such as OA status, publication date, publisher, and concepts, among others, for each DOI. After matching, we obtained article information from OpenAlex for 2,154,524 journal articles.

We considered five categories for the OA policy in our study, following the classification scheme proposed by \cite{piwowar_state_2018}:
\begin{enumerate}
\item Gold: Refers to articles published in an open-access journal that is indexed by the Directory of open access Journals (DOAJ).
\item Green: Refers to articles that are available on the publisher's webpage but are also freely accessible in a pre-print repository.
\item Hybrid: Refers to articles that are available for free under an open license in a toll-access journal.
\item Bronze: Refers to articles that can be read for free on the publisher's webpage but do not have an easily identifiable license.
\item Closed: Refers to all other articles, including those shared only on an academic social network or in Sci-Hub.
\end{enumerate}

Additionally, we collected journal information for conducting a regression analysis on the influence of OA. We obtained this information by downloading data from Scimago\footnote{\url{https://www.scimagojr.com/aboutus.php}.}. Scimago is an open-access resource that provides an internationally accepted journal rank indicator for analysis in the fields of scientometrics and infometrics~\cite{falagas_comparison_2008, yuen_comparison_2018, gonzalez-pereira_new_2010}. We equipped each journal with an SJR score, H index, and other relevant information.

\subsection{Model Specification}

\textbf{Dependent variable}\\
To assess the potential advantage of OA articles in Wikipedia, we defined a binary dependent variable, denoted by \textit{\textbf{is\_wiki}}, that indicates whether an article has been cited in Wikipedia or not. Since our primary dataset consists solely of articles cited in Wikipedia, we use OpenAlex to obtain negative samples of articles not cited in Wikipedia, via stratified sampling.
\\
\textbf{Independent variable}\\
To assess the impact of OA articles on their citation rates in Wikipedia, we analyze two types of variables: article-level and journal-level. At the article level, we consider the number of citations (\textit{\textbf{times\_cited}}), whether the article is OA or not (\textit{\textbf{is\_oa}}), the time of publication (\textit{\textbf{article\_age}}), and the field of research (\textit{\textbf{concept}}). These features have been shown to have an influence on citation impact in previous studies~\cite{colavizza_citation_2020, gargouri_self-selected_2010, yegros-yegros_does_2015, struck_modelling_2018, teplitskiy_amplifying_2017, nielsen_scientific_2007}. At the journal level, we primarily consider the Scimago journal rank (\textit{\textbf{SJR}}). To accurately represent the Scimago journal rank for each article, we use the Scimago journal rank assigned to journals for the same year in which an article was published. Since a year-by-year breakdown for journal ranks was only available from 1999-2020, we assign the journal rank corresponding to the year 1999 for those articles published before 1999, as it is the earliest representation of a journal’s rank. This range (i.e. published before 1999) accounts for 29\% of citations from our curated set of citations in Wikipedia.

Although these variables have been widely used to model citation impact in previous studies, little analysis has directly linked these indicators to whether an article is cited in Wikipedia or not, specifically for different OA policies.

\begin{figure}[H]
\centering
\includegraphics[width=.65\linewidth]{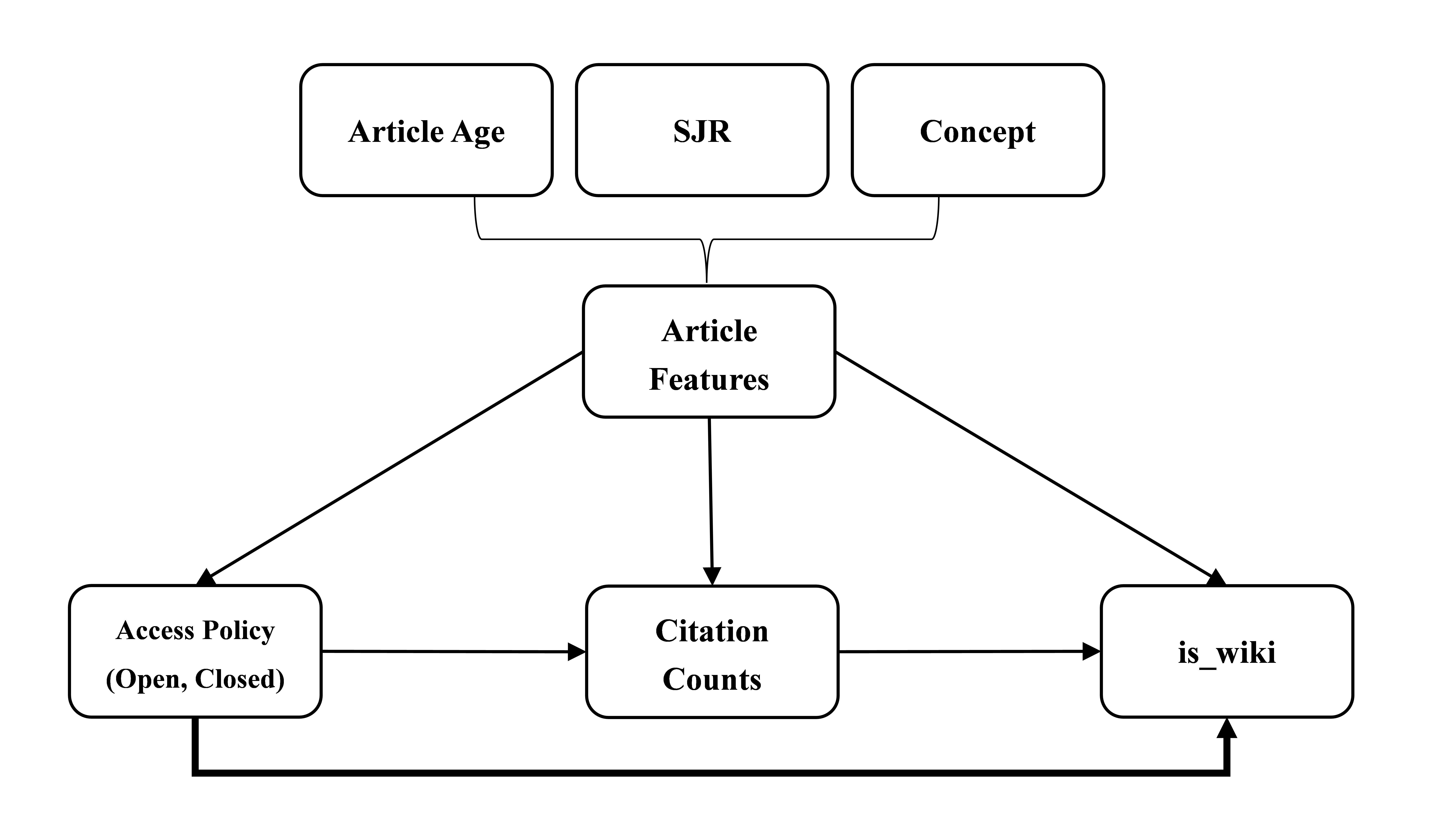}
\caption{Assumed causal structure of Wikipedia's OA citation adoption effect, with a black line representing an assumed causal relationship between two variables.}
\label{causality}
\end{figure}

In this study, we use logistic regression as our model, which is usually used to analyze the relationship between a binary dependent variable and one or more independent variables. The logistic regression weights represent the size of the individual contributions of each predictor variable to the target variable. Figure \ref{causality} illustrates the assumed causal structure of Wikipedia's OA citation adoption effect, with a black line representing an assumed causal relationship between two variables. Specifically, we assume that the likelihood of a journal article being cited in Wikipedia is directly influenced by its article features, citation counts, and OA policy. At the same time, the OA policy can also influence the citation counts of the article, causing a further mediated effect on the adoption of this article in Wikipedia. With our models, we are interested in measuring both the (controlled) direct effect and the total effect of OA policy on being cited in Wikipedia. The former is shown as a black thick line in Figure \ref{causality}, while the latter accounts for both the direct and the mediated (via citation counts) effects.

\subsection{Dataset construction}
We aim to create a balanced dataset of journal articles that can be analyzed using regression analysis. The initial dataset was sourced from Wikipedia and contained 1,499,021 unique scientific articles. To initiate our regression analysis, we constructed the dataset which was constructed by adding Journal, Year of publication and Concept as stratifying variables, and it will be used to account for the influence of \textit{\textbf{concept}}. To restrict ourselves to root-level concepts and avoid any ambiguity, we filtered the citations to only include those with one associated concept, resulting in a set of 410,573 articles. Subsequently, we assemble corresponding sets of articles for these two datasets from OpenAlex based on the stratifying variables and excluding those already cited in Wikipedia. To reduce noise in the sampling strategy, we removed journals with no corresponding name available in Scimago and those with less than 20 citations. We also removed all articles published before 1900 to remove sparsely mentioned dates and accept a slight recency bias.

After pre-processing, we group the number of articles within each stratum and proceed as follows:
\begin{enumerate}
\item Filter the whole set of OpenAlex articles to those matching the fields in the strata.
\item If there are fewer articles than the curated Wikipedia dataset in this filtered set, discard the strata and remove the respective articles from the curated dataset.
\item Otherwise, randomly sample the same number of articles from the filtered set and add it to the set of negative samples.
\end{enumerate}

After iterating through all strata (90,019 strata), we derived a final negative set comprising 261,230 entries. When combined with the corresponding sets of Wikipedia-cited articles, this yields a comprehensive dataset totalling 522,460 entries.

To ensure the robustness of our sampling methodology, we repeated the process five times, resulting in five different sets that were used in the analyses. Although our method of matching strata to construct a set of negative samples is an approximation of the more rigorous method of propensity score matching (PSM), the discrete nature of our strata and the large population size contribute to the robustness of our analysis. A descriptive overview of this curated dataset can be found in Tables \ref{tab:table15}, \ref{tab:table16} and \ref{tab:table11} in the appendix.

\section{Results}
We have augmented our analysis by incorporating additional metadata from OpenAlex and Scimago, which allowed us to obtain information for 98.0\% (2,154,524) of the total 2,197,461 citations to a valid DOI. From these, we extracted 1,499,021 publications (unique DOIs) and an associated open access (OA) status. Our findings show that 46.5\% (1,021,820 out of 2,197,461) of the citations and 44.1\% (661,068 out of 1,499,021) publications were OA (i.e., not closed). 

\subsection{Characterizing open access Articles within Wikipedia}
\begin{figure}[H]
\begin{subfigure}[b]{.5\textwidth}
  \centering
  \includegraphics[width=.99\linewidth]{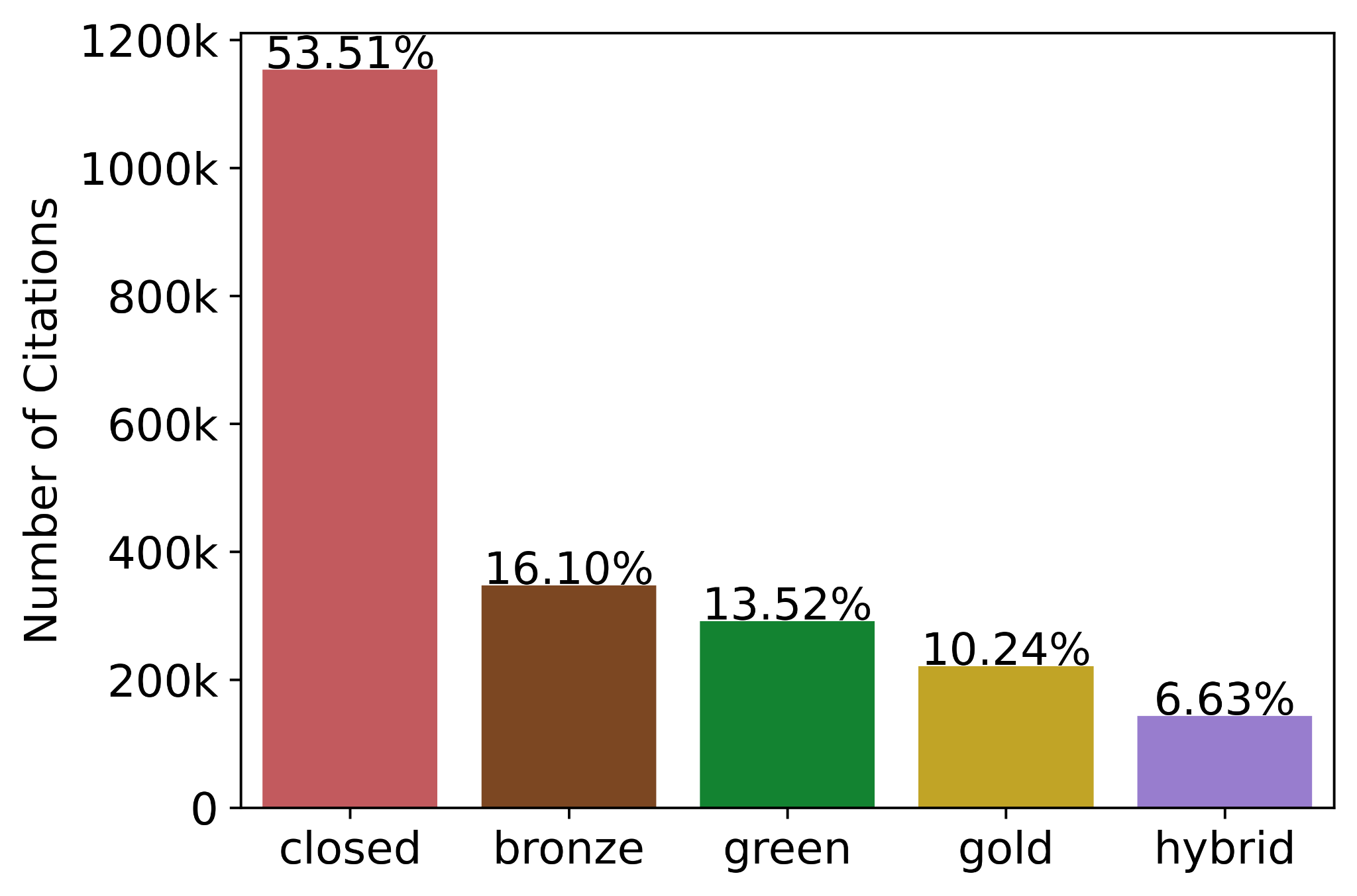}
  \caption{Over citations.}
  \label{fig:Figure1_1}
\end{subfigure}%
\hfill
\begin{subfigure}[b]{.5\textwidth}
  \centering
  \includegraphics[width=.99\linewidth]{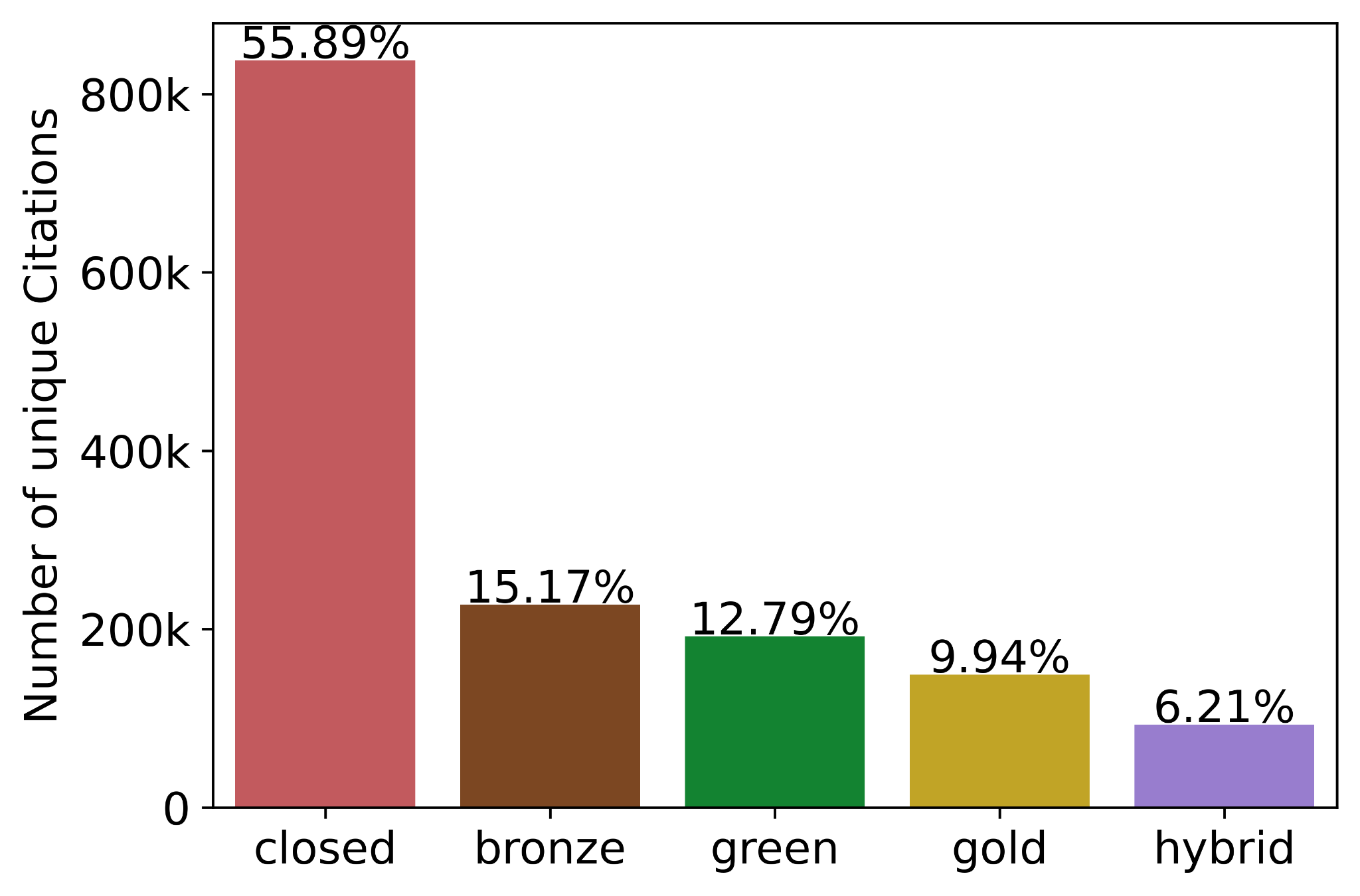}
  \caption{Over unique journal articles.}
  \label{fig:Figure1_2}
\end{subfigure}
\caption{Distribution of open access status by policy.}
\label{fig:Figure1}
\end{figure}

We present our findings on the distribution of open access (OA) policies in Wikipedia citations in figure~\ref{fig:Figure1}. Our results demonstrate that the most commonly observed OA policy in Wikipedia citations is the bronze policy, which is consistent with trends in scholarly literature~\cite{piwowar_state_2018}. The second most popular OA policy observed in Wikipedia citations is Green, which is significantly more prevalent than the Gold policy. 

\begin{figure}[htbp]
\centering
\includegraphics[width=.7\linewidth]{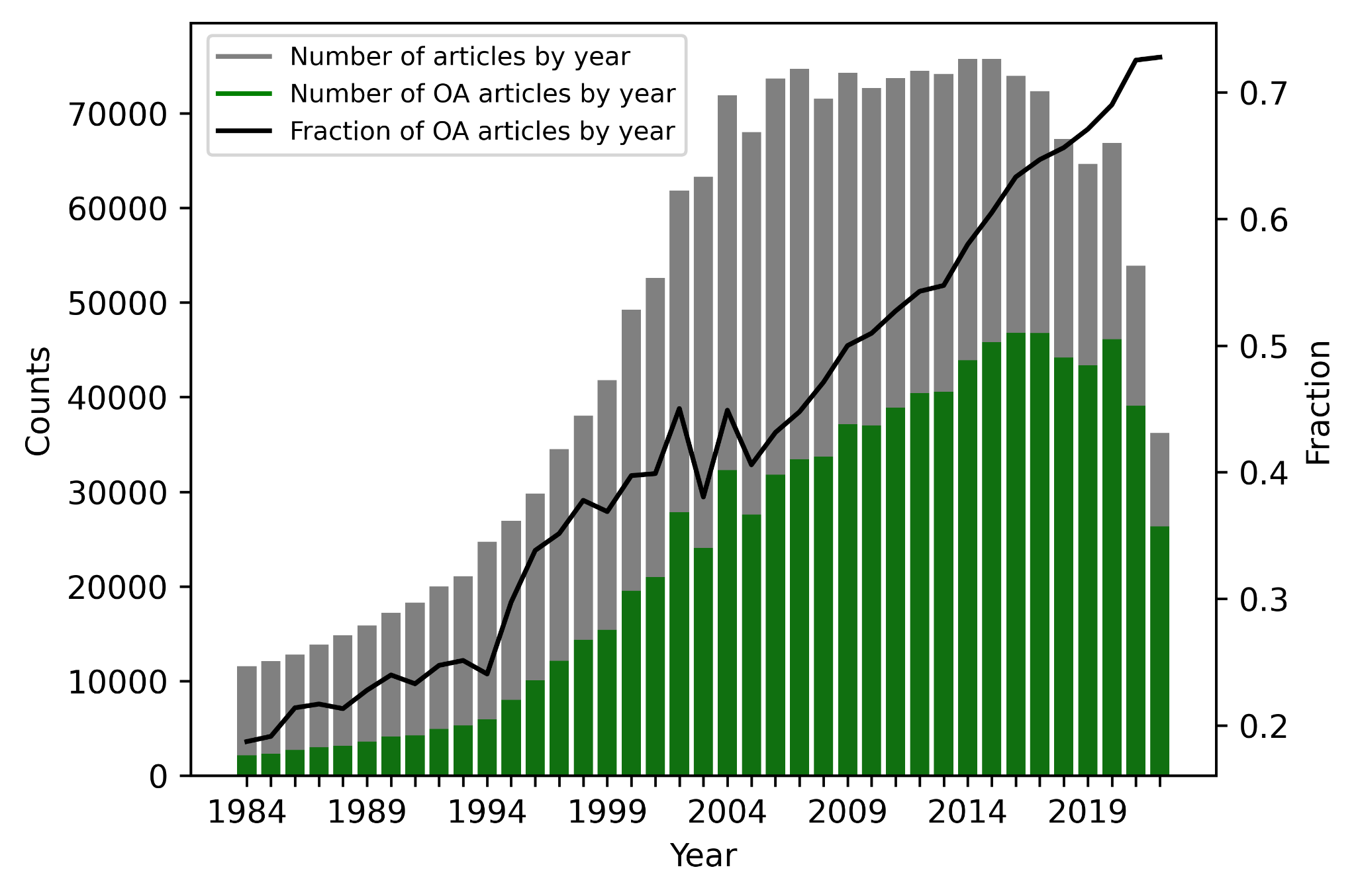}
\caption{Fraction of OA citations by publication date of citation. 
The grey bars represent the number of scientific articles cited by Wikipedia, grouped by their publication year. The green bars denote the number of OA articles among these cited articles. Both sets of bars utilize the left y-axis (Number of articles). The black line illustrates the percentage of OA articles concerning their publication year, calculated as the ratio of the green bar (Number of OA articles) to the grey bar (number of total articles). The black line employs the right y-axis (Fraction of OA).}
\label{figure4}
\end{figure}

Figure \ref{figure4} displays OA distribution by publication date of citations. The grey bar denotes the count of articles cited by Wikipedia in each publication year (including both OA and non-OA articles), the green bar signifies the number of OA articles in each grey bar, and the black line illustrates the proportion of OA articles cited in each corresponding publication year. The left y-axis corresponds to the article count, while the right y-axis indicates the fraction of OA articles. Overall, there has been a consistent rise in the proportion of citing Open Access (OA) articles over the last four decades. This trend suggests a growing prevalence of OA articles, hinting at their potential influence on the trajectory of scientific representation in Wikipedia. Specifically, among the recently published scientific articles, Wikipedia appears to show a preference for citing Open Access (OA) articles in its citations. Notably, the percentage of OA article citations surpassed 50\% after the year 2015.

We proceeded to examine the breakdown of OA status and OA policies by journals in our dataset, which includes 40,806 journals. In order to visualize this information effectively, we determined the number of citations for each journal and selected the top 20 for analysis. Figure \ref{figure7} displays the total number of citations for the top 20 journals, with blue and orange representing OA and non-OA articles, respectively. As found in previous studies, some high-impact journals such as \emph{Nature}, \emph{PNAS}, and \emph{Science} appear frequently on Wikipedia~\cite{nielsen_scientific_2007} and account for 5.7\% of all citations. However, inferring article OA policy based on whether journals are classified as `Open Access' or `Closed Access' can be misleading~\cite{teplitskiy_amplifying_2017}, as there is a high variance in OA status among articles within journals. For example, while some articles in \emph{Nature} and \emph{Science} are OA, there are also non-OA articles. Therefore, it is inappropriate to study the relationship between open access and Wikipedia based on the journal level. 

To further investigate the distribution of OA policies among the top 20 journals, we plotted the data in Figure \ref{figure8}. Our analysis revealed a growing trend of bronze OA policies among journals. However, some journals that classify themselves as OA, such as ``\emph{Journal of Biological Chemistry}"\footnote{\url{https://www.elsevier.com/journals/journal-of-biological-chemistry/0021-9258/open-access-journal}.} and ``\emph{PLOS one}"\footnote{\url{https://journals.plos.org/plosone/s/journal-information\#loc-open-access}.}, have a significant proportion of articles classified as Hybrid or Gold OA. While there may be limitations in the classification of OA articles by OpenAlex, we accepted their classifications in our study.

\begin{figure}[ht]
\centering
\includegraphics[width=.7\linewidth]{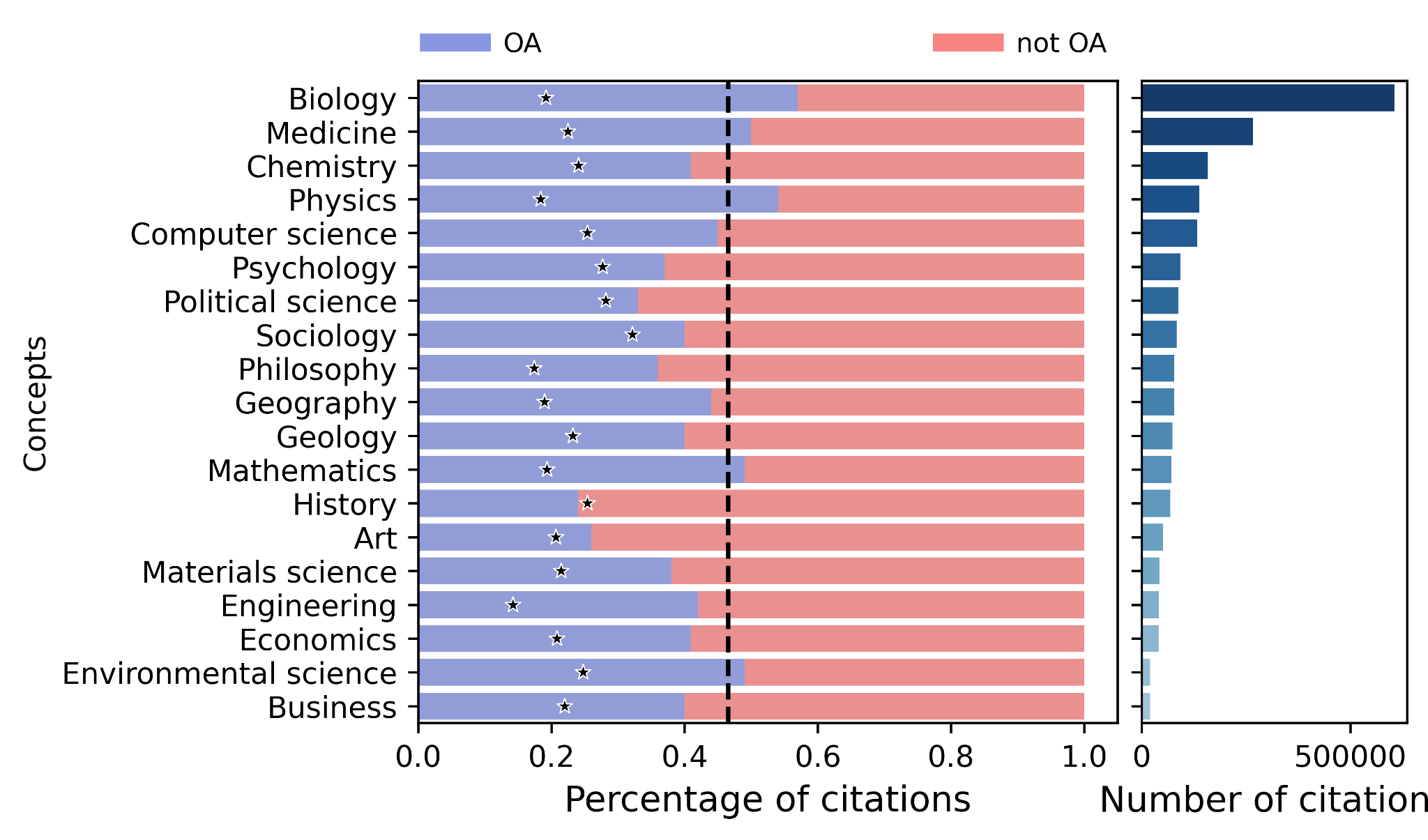}
\caption{Distribution of OA status and count of citations by OpenAlex concept.
The blue bars depict the fraction of OA citations within each OpenAlex concept, while the red bars represent the fraction of closed articles. The black dotted line indicates the overall percentage of OA articles in our Wikipedia dataset, and the black star signifies the percentage of OA articles by concepts in the entire OpenAlex dataset. The right chart displays the number of citations for each concept in Wikipedia, with the concepts arranged in order based on the number of citations.}
\label{figure2}
\end{figure}

Furthermore, we present an analysis of the distribution of OA status by OpenAlex concepts, as shown in Figure \ref{figure2} and Figure \ref{figure3}. We used OpenAlex, a dataset containing 65k concepts and 19 root-level concepts. We employed fractional counting to determine the number of citations for each root-level concept. Given that 46.5\% of citations on Wikipedia are OA, we used this percentage as a baseline for OA proportionality, represented by the black dotted line in Figure \ref{figure2}. Additionally, we included the percentage of OA articles for each concept in the entire OpenAlex dataset, represented by the black star, as a reference for the broader scientific landscape. Our analysis revealed substantial variance in field-specific OA proportions.

In Figure \ref{figure2}, the left part shows the percentage of citations with OA status for each concept, and the right part represents the total number of citations of corresponding concepts, arranged from the largest to the smallest. 
Remarkably, the OA proportions for all concepts in Wikipedia significantly exceed those observed from the perspective of OpenAlex. This reaffirms the pivotal role of OA articles in shaping the information sources within Wikipedia. From the perspective of Wikipedia, while the average Open Access (OA) proportion stands at 46.5\%, there are still 5 concepts that surpass this average: Biology (57\%), Physics (53\%), Medicine (50\%), Environment science (49\%) and Mathematics (49\%). In contrast, Political science (33\%), Art (27\%), and History (24\%) displayed the lowest proportions of OA articles among the cited Wikipedia articles. In general, compared to the humanities, Wikipedia not only exhibits more citations of scientific articles on STEM-related concepts but also relies on OA articles.

\subsection{OA Citation Advantage}
To comprehensively assess the impact of an article's OA status on its likelihood of being cited by Wikipedia, we designed a series of statistical models leveraging the datasets outlined in the data and methods section. The objective of this analysis is to elucidate the role of OA articles within the scientific discourse of Wikipedia, aiming to discern any potential advantages associated with their citation patterns.

\subsubsection*{Model results}
We use logistic regression for interpretability and expressiveness, and we also use log transforms on the continuous variables.
To comprehensively evaluate the overall impact of OA status on citation adoption, our principal logistic regression model, designed to scrutinize the influence of $\textit{is\_oa}$, is formulated as follows:

\begin{align}
    is\_wiki =  is\_oa + ln(article\_age) + ln(SJR) + concept +is\_oa*ln\_article\_age
\end{align}

To examine the direct influence of OA status on the adoption of citations in Wikipedia, considering the interplay between citation count and open access policy, we introduce the second formula:

\begin{align}
    is\_wiki =  & is\_oa + ln(article\_age) + ln(SJR) + concept +ln(times\_cited + 1) \\
    & +is\_oa*ln\_article\_age \notag
\end{align}

To validate the robustness of our model, we assess the statistical significance of each coefficient across all five samples. A coefficient is deemed statistically insignificant if it lacks significance in the results of at least one sample. Subsequently, we present the effects in terms of odds ratios, calculated based on the mean odds ratios derived from the outcomes across all five samples. In addition, we conducted a multicollinearity check on the variables in the model and found that all Variance Inflation Factors (VIF) were below 10, indicating the absence of significant multicollinearity issues.

Our analysis of the data is presented in Table \ref{tab:table1_new}, which lists the regression results of model 1 and model 2. From Table \ref{tab:table1_new}, it is evident that OA articles exhibit substantially higher odds of being cited in Wikipedia compared to closed-access articles. Specifically, in model 1, OA articles have 80\% higher odds of being cited, and this becomes 64.7\% in model 2 when considering citation counts. These findings underscore the critical role of OA status in the utilization of scientific articles as references on Wikipedia.

Moreover, incorporating citation counts into the model enhances the interpretability and reveals additional insights. Similar to OA status, citation counts play a significantly positive role in the odds of scientific articles being cited in Wikipedia. This suggests that articles with higher citation counts are more likely to be referenced on Wikipedia, reflecting their impact and visibility within the scientific community. In terms of conceptual classification analysis, we chose biology, which has the highest citation count on Wikipedia, as our reference point. By incorporating citation counts into our analysis, we identified 14 concepts that significantly influence the likelihood of articles categorized as open access (OA) being cited on Wikipedia. Notably, several concepts from the humanities and social sciences, such as Art, History, Philosophy, and Political Science, exhibited higher notably positive coefficients. This finding reflects the unique characteristics of these domains, known for being low-citation fields~\cite{patience_citation_2017}, despite their substantial importance within the Wikipedia ecosystem. Additionally, Environmental Science also demonstrated a high coefficient, likely due to its interdisciplinary nature, which incorporates knowledge from both natural and social sciences.

Furthermore, the age of the article demonstrates a modest yet significantly negative effect on the likelihood of OA articles being cited by Wikipedia. This finding implies that newer publications are more likely to be cited by Wikipedia compared to older articles, indicating a preference for recent and up-to-date scientific content on the platform.

Despite the SJR demonstrating a significantly negative effect, insights can be gleaned from the distribution of SJR among Wikipedia citations, as depicted in Figure \ref{figure_SJR}. In Figure \ref{figure_SJR}, the x-axis represents the SJR value obtained from Scimago, while the y-axis represents the proportion of Wikipedia citations. It is evident that nearly 90\% of the cited journals on Wikipedia have a small SJR, which is less than 10. Additionally, the mean SJR in our dataset is 3.68, with a third quartile of 4.38. This finding further supports our regression results, indicating that in Wikipedia, most citations come from journals with small SJR values. Thus, as the SJR decreases, the likelihood of OA articles being cited by Wikipedia increases.

\begin{table}[htbp]
    \centering
    \caption{Regression results for the first sample with models 1 and 2.}
    \label{tab:table1_new}
    \begin{tabular}{|l|c|c|c|c|c|c|}
        \hline
        Regression Model & 
        \multicolumn{3}{c|}{Model 1 ($R^{2}=0.00034$)} & 
        \multicolumn{3}{c|}{Model 2 ($R^{2}=0.07182$)} \\
        \hline
        Feature                 & Coef   & Odds Ratios & P\textgreater{}z & Coef   & Odds Ratios & P\textgreater{}z \\
        \hline
        Intercept               & -0.363 & 0.695       & 0       & -0.979 & 0.376       & 0       \\
        ln1p\_times\_cited      &        &             &         & 0.442  & 1.557       & 0       \\
        ln(article\_age)         & 0.064  & 1.066       & 0       & -0.067 & 0.935       & 0       \\
        ln(SJR)                 & -0.002 & 0.998       & 0.487   & -0.246 & 0.782       & 0       \\
        is\_oa                  & 0.588  & 1.800       & 0       & 0.499  & 1.647       & 0       \\
        is\_oa:ln\_article\_age & -0.091 & 0.900       & 0       & -0.103 & 0.902       & 0       \\
        Art                     & -0.001 & 0.999       & 0.980   & 0.669  & 1.952       & 0       \\
        Business                & 0.001  & 1.001       & 0.992   & 0.38   & 1.462       & 0       \\
        Chemistry               & -0.002 & 0.998       & 0.845   & 0.014  & 1.014       & 0.237   \\
        Computer science        & -0.004 & 0.996       & 0.846   & 0.251  & 1.285       & 0       \\
        Economics               & -0.006 & 0.994       & 0.885   & 0.042  & 1.043       & 0.387   \\
        Engineering             & 0.000  & 1.000       & 0.999   & 0.523  & 1.688       & 0       \\
        Environmental science   & 0.002  & 1.002       & 0.987   & 0.75   & 2.118       & 0       \\
        Geography               & -0.007 & 0.993       & 0.877   & 0.583  & 1.791       & 0       \\
        Geology                 & 0.008  & 1.008       & 0.626   & 0.09   & 1.095       & 0       \\
        History                 & -0.004 & 0.996       & 0.914   & 0.646  & 1.908       & 0       \\
        Materials science       & 0.009  & 1.009       & 0.822   & -0.069 & 0.934       & 0.109   \\
        Mathematics             & 0.002  & 1.002       & 0.931   & 0.415  & 1.514       & 0       \\
        Medicine                & 0.003  & 1.003       & 0.675   & 0.03   & 1.03        & 0       \\
        Philosophy              & -0.003 & 0.997       & 0.915   & 0.703  & 2.02        & 0       \\
        Physics                 & -0.012 & 0.988       & 0.386   & 0.227  & 1.255       & 0       \\
        Political science       & -0.006 & 0.994       & 0.851   & 0.681  & 1.977       & 0       \\
        Psychology              & -0.002 & 0.998       & 0.905   & -0.069 & 0.934       & 0       \\
        Sociology               & 0.002  & 1.002       & 0.981   & 0.466  & 1.594       & 0       \\
        \hline
    \end{tabular}
\end{table}

To gain insight into the interaction between open access (OA) status and citation counts in Wikipedia, we use formula 2 to create a graph plotting the functions that contain these two variables and article features. 

\begin{figure}[htbp]
\centering
\includegraphics[width=.7\linewidth]{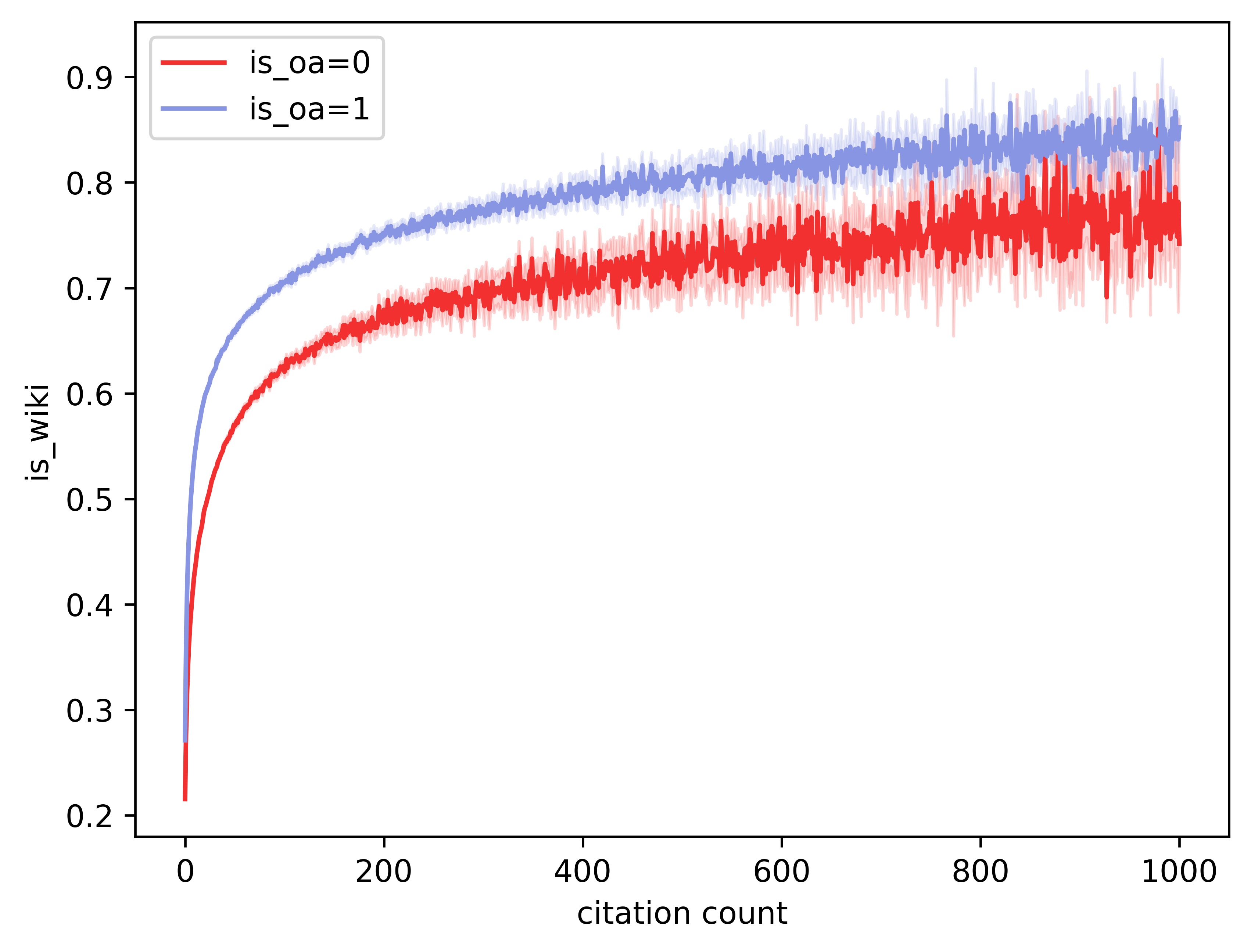}
\caption{OA adoption effect at varying citation counts, based on model 2.}
\label{fig6:OA adoption effect by citation counts}
\end{figure}

The graph, shown in Figure \ref{fig6:OA adoption effect by citation counts}, displays the dependent variable, \textit{\textbf{is\_wiki}}, on the y-axis and the citation counts (variable \textit{\textbf{times\_cited}}) on the x-axis. Articles are grouped according to their open access (OA) status. We plot the average model prediction calculated on each group using the first data sample and provide 95\% bootstrapped confidence intervals for each group (faded color). The red line illustrates the trend of OA adoption by citation count under the condition that the OA status is closed, while the blue line shows the trend under the condition that the OA status is open. This graph reveals several insights. Firstly, when the citation counts are very low (near 0), there is a significant initial citation advantage for OA articles compared to the closed articles. As the citation count increases (up to 200), this advantage gradually expands. However, as the citation count continues to grow, this advantage becomes less distinguishable. In our previous work~\cite{yang2021map}, we show that articles cited fewer than 100 times account for 70\% of the total cited articles, and only about 3\% of articles are cited 1,000 times or more. Therefore, most citations in Wikipedia benefit from this OA effect. 
Our speculation regarding the OA adoption effect is that Wikipedia editors might find it easier to discover and access open research results earlier in the publication timeline before these articles accumulate citations and gain peer recognition.

\begin{figure}[H]
\centering
\includegraphics[width=.7\linewidth]{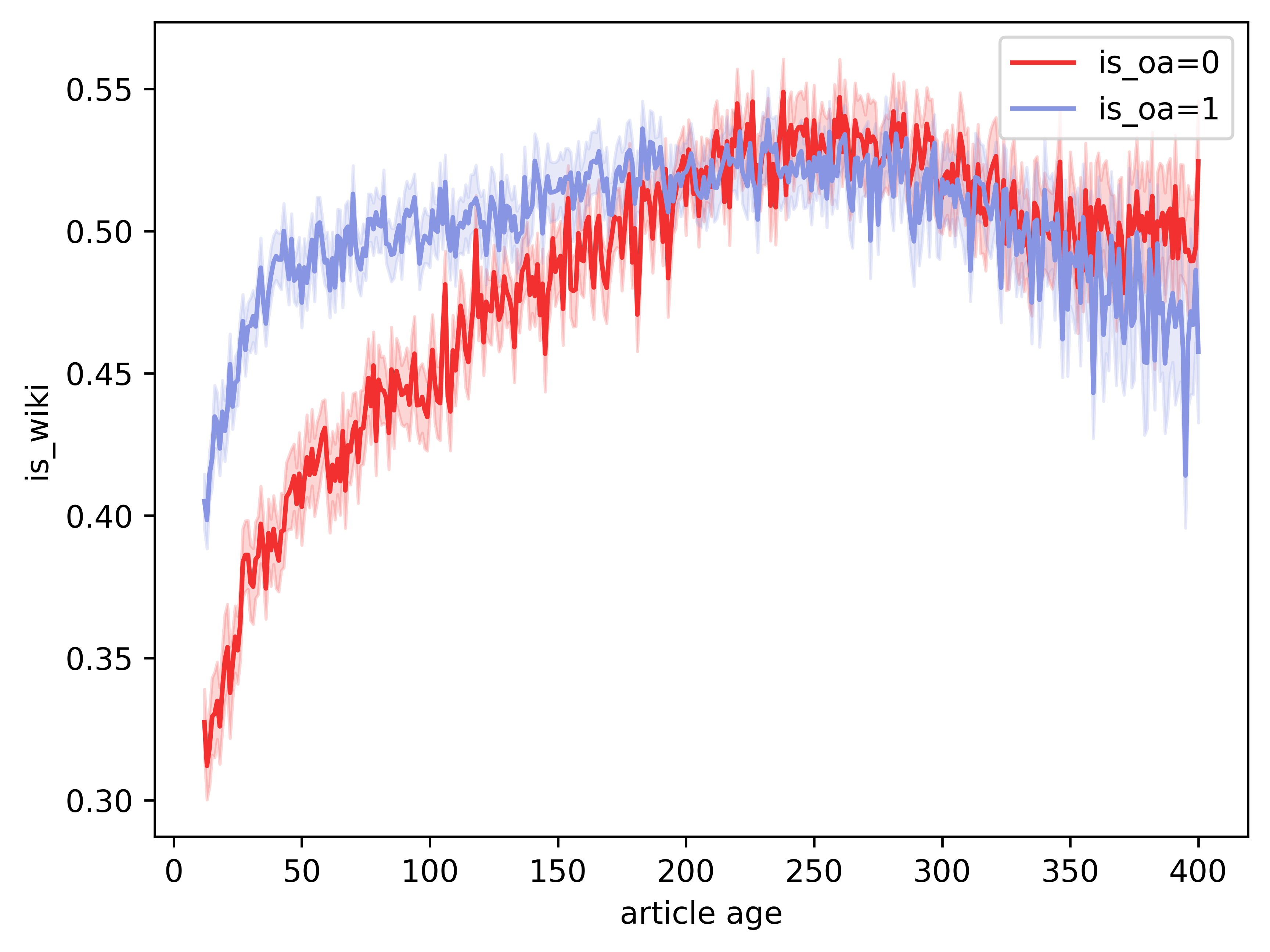}
\caption{OA adoption effect at varying article age, based on model 2.}
\label{fig:OA adoption effect by article age}
\end{figure}

In addition, we examined the interaction between OA status and article age on Wikipedia using model 2, as depicted in Figure \ref{fig:OA adoption effect by article age}. The figure highlights a significant advantage for younger articles, particularly those with an age of less than 200 months (16.6 years). Here, an OA article demonstrates a 10\% higher likelihood of being cited by Wikipedia compared to a closed-access article. However, as the article age increases to around 240 months (or 20 years), the odds of adoption begin to decline. This observation underscores Wikipedia's preference for newer articles over older ones, particularly those published within the last four years.

Moreover, we employed two regression models to investigate the impact of OA policy on citation adoption, using 'closed' as the baseline. The results, presented in Table \ref{tab:table5}, demonstrate that all OA policies significantly enhance the overall adoption effect for OA articles. Additionally, the interaction between OA policies and article age exhibits a significant negative effect on OA adoption. Table \ref{tab:table6} presents the results using the second model to explore the indirect effect of OA policy, revealing a similar trend, although the bronze policy exhibits a slightly significant negative impact. To validate the robustness of our findings, we conducted regressions across all five samples, and the results are reported in Tables \ref{tab:table5_all} and \ref{tab:table6_all}.

\begin{table}[htb]
\centering
\caption{Coefficients for OA adoption by policy. Results for the first sample, model 1, $R^{2}=0.0013$.}
\label{tab:table5}
\begin{tabular}{|l|c|c|c|}
\hline
Feature & coef & odds\_ratios & P\textgreater{}z \\ \hline
bronze                       & 0.101   & 1.106 & 0.000 \\ \hline
gold                         & 0.032   & 1.032 & 0.000 \\ \hline
green                        & 0.162   & 1.176 & 0.000 \\ \hline
hybrid                       & 0.190   & 1.210 & 0.000 \\ \hline
ln\_article\_age & 0.019   & 1.019 & 0.000 \\ \hline
ln(SJR)       & -0.036  & 0.965 & 0.000 \\ \hline
bronze:ln\_article\_age     & -0.0398 & 0.961 & 0.000 \\ \hline
gold:ln\_article\_age        & -0.1392 & 0.870 & 0.000 \\ \hline
green:ln\_article\_age       & -0.1868 & 0.830 & 0.000 \\ \hline
hybrid:ln\_article\_age      & -0.2230 & 0.800 & 0.000 \\ \hline
\end{tabular}
\end{table}

\begin{table}[htb]
\centering
\caption{Coefficients for OA adoption by policy. Results for the first sample, model 2, $R^{2}=0.073$.}
\label{tab:table6}
\begin{tabular}{|l|c|c|c|}
\hline
Feature &coef &odds\_ratios &P\textgreater{}z \\ \hline
bronze                         & -0.111                        & 0.895                        & 0.054 \\ \hline
gold                           & 1.090                         & 2.974                        & 0.000 \\ \hline
green                          & 0.834                         & 2.302                        & 0.000 \\ \hline
hybrid                         & 1.156                         & 3.177                        & 0.000 \\ \hline
ln\_article\_age   & -0.070 &  0.933 & 0.000 \\ \hline
ln(SJR)         &  -0.244 &  0.783 & 0.000 \\ \hline
bronze:ln(article\_age)        &  0.029  &  1.030 & 0.007 \\ \hline
gold:ln\_article\_age         &  -0.245 &  0.783 & 0.000 \\ \hline
green:ln\_article\_age         &  -0.171 &  0.843 & 0.000 \\ \hline
hybrid:ln\_article\_age        &  -0.274 &  0.760 & 0.000 \\ \hline
$\ln(1+\textit{times\_cited})$ &  0.447  &  1.564 & 0.000 \\ \hline
\end{tabular}%
\end{table}

\section{Discussion}

The surge in popularity and growth of Open Access (OA) has significantly contributed to the dissemination of scientific knowledge. Our research underscores Wikipedia's growing reliance on OA articles, constituting 46.5\% of all scientific citations on the platform, 
a notable increase from the 31.2\% reported in the prior study~\cite{pooladian2017methodological}. This trend has witnessed continuous growth, notably seen in scientific articles cited by Wikipedia which were published after 2011, where a minimum of 50\% are open access. These findings align with the broader scientific community's trend, as evidenced by the percentage of OA articles steadily increasing to 28\% in 2018, with OpenAlex reporting 47\%~\cite{piwowar_state_2018}. Despite high-impact journals remaining a preferred source for Wikipedia~\cite{nielsen_scientific_2007}, variations in the distribution of OA articles within journals emphasize the necessity for a nuanced, article-level approach.

Our examination of OA policies in scientific articles and Wikipedia unveils a parallel trend~\cite{piwowar_state_2018}. Bronze policy (16.10\%) and green policy (13.52\%) dominate as the most common OA policies in Wikipedia. Notably, the prevalence of Green policy in Wikipedia surpasses that in scientific articles, suggesting distinctions arising from differences in reference acquisition methods between Wikipedia editors and researchers. This trend further reinforces the importance of not overlooking articles accessible through green routes, thereby avoiding underestimating the impact open access can have on disseminating scientific knowledge through Wikipedia~\cite{elsabry2017needs}.

Our study further reveals disparities in OA Wikipedia citations across disciplines, with biology, physics, and mathematics exhibiting higher OA citation rates, while social sciences and humanities show comparatively lower rates. Nevertheless, Wikipedia's robust reliance on OA articles persists across all OpenAlex root concepts.

The odds of an OA article being cited in Wikipedia over a closed-access article increase by an average of 64.7\%. Moreover, the likelihood of OA articles being cited increases with article age and citation count. Despite the significantly negative effect of SJR, an examination of SJR distribution among Wikipedia citations reveals that over 90\% have an SJR lower than 10, with nearly 80\% below 5. Analysis of citation counts and article age suggests that OA articles with higher citation counts are more likely to be referenced in Wikipedia, indicating the preference for reliable sources by Wikipedia editors. Furthermore, Wikipedia editors demonstrate a capacity to rapidly update scientific knowledge, particularly focusing on articles published within the past four years, reflecting responsiveness to new scientific developments.

We acknowledge certain limitations in our study. For instance, the exclusion of conference papers and earlier literature due to our focus on articles with DOIs suggests that future research could consider additional sources. While our regression model accounted for significant factors, such as OA status, OA policy, and citation counts, other causal variables like article length may influence article citations on Wikipedia. Additionally, our study did not consider time as an analytical dimension, prompting future research to delve into Wikipedia's edit history for specific data at the time of article citation, facilitating a deeper understanding of the causal mechanisms underpinning the interplay between open access and Wikipedia.

\section{Conclusion}

This study examined the impact of open access (OA) on Wikipedia by analyzing article-level features. By utilizing a large dataset of citations from Wikipedia, coupled with OA-related metrics from OpenAlex and journal information from Scimago, we investigated the prevalence and significance of OA in Wikipedia. The results show that OA plays a crucial role in Wikipedia: OA articles are increasingly more cited over time, and have a higher chance of being cited in Wikipedia than similar closed-access articles. In particular, articles with high citation counts and published within about 4 years are substantially more likely to be cited in Wikipedia. These findings underscore the effective role of open access in disseminating scientific knowledge to a broader audience, particularly through important platforms like Wikipedia. Additionally, newer and more influential articles are more likely to gain visibility and attention from the public.

Our study provides a foundation for further research on Wikipedia and open science more broadly. Future studies should broaden source and variable coverage to better unpack the OA effect on Wikipedia. Furthermore, other forms of open science could be analyzed using a similar lens, for example, open research data and software. In conclusion, this study sheds light on the significance of OA in Wikipedia and potentially its broader impact. We believe that our findings will serve as a starting point for further research and contribute to the understanding of the impact and dissemination of OA.

\section{Code and data availability statement}
The code to replicate our work is made available online: \url{https://github.com/alsowbdxa/Open_access_and_wikipedia}. The Wikipedia Citations dataset is openly available~\cite{kokash_2024_10782978}, while access to OpenAlex can be requested through their portal. All other supporting datasets we used are openly available and referenced from the Data and Methods section.  

\section{Appendix}

Our Appendix comprises three subsections: Figures, Tables, and Regression results supplements. Although these results offer a more extensive understanding of our research, they do not constitute the principal outcomes. Therefore, we have relocated them to the appendix for further reference.

\subsection{Figures}

Presented below are two figures depicting the distribution of OA status and policies among the top 20 journals. We have discussed it in the results part. This observation underscores the significance of conducting an article-level analysis for a more comprehensive understanding of the subject matter.

\begin{figure}[H]
\centering
\includegraphics[width=.7\linewidth]{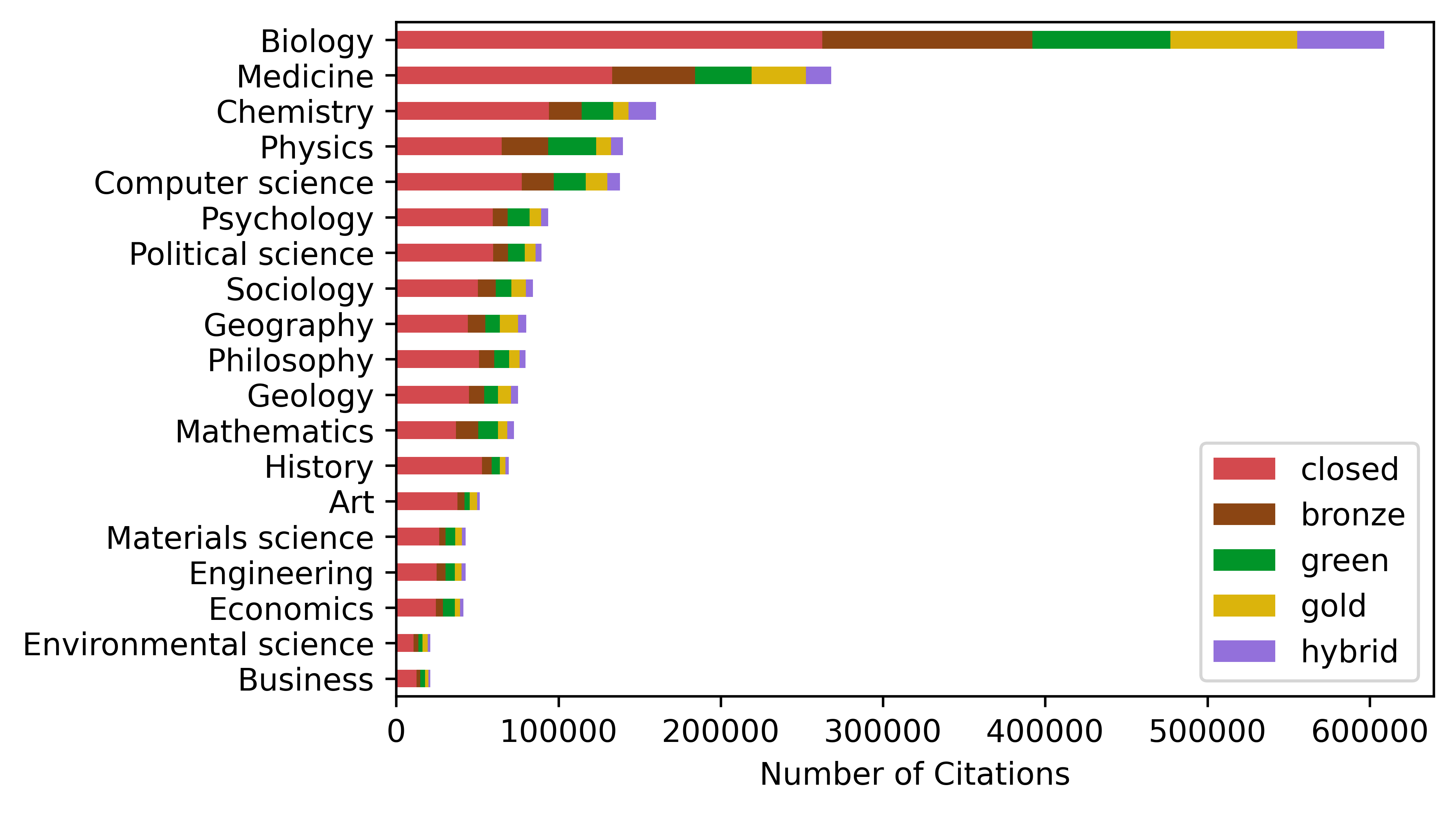}
\caption{Distribution of OA policies by OpenAlex concept.}
\label{figure3}
\end{figure}

In Figure \ref{figure3}, we illustrate the distribution of OA policies across various concepts. Our analysis reveals that bronze and green policies predominantly characterize most concepts in OA articles, except for Art, where the gold policy assumes significance.

\begin{figure}[H]
\centering
\includegraphics[width=.7\linewidth]{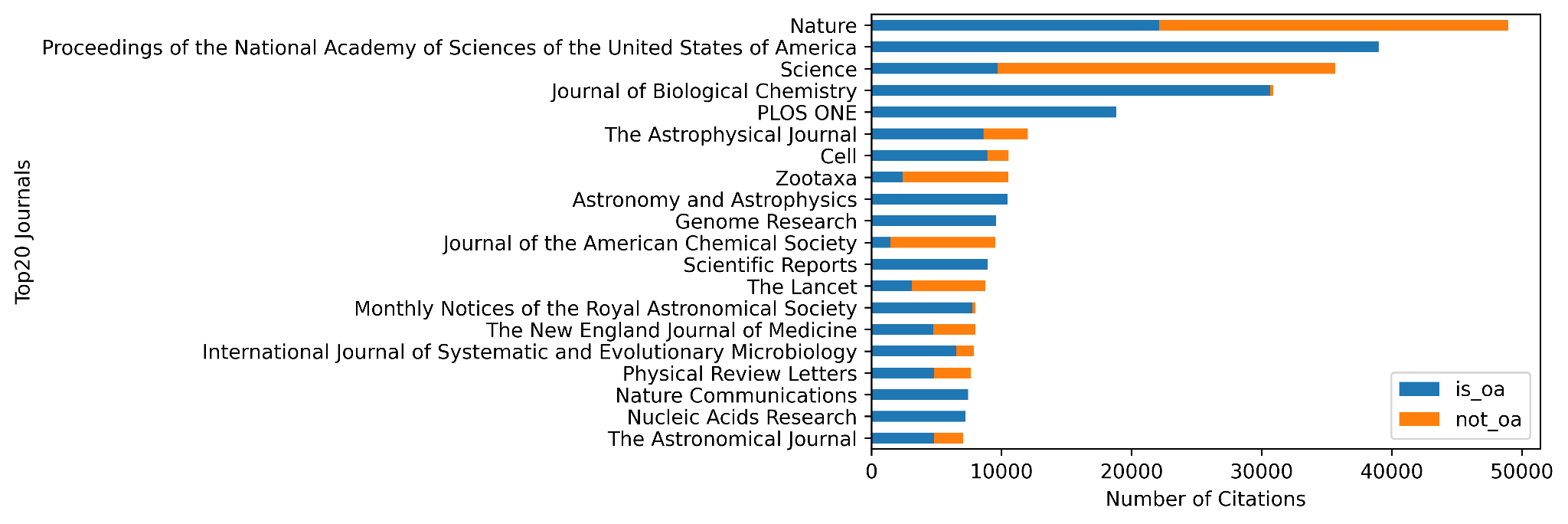}
\caption{Distribution of OA status by top 20 journals.}
\label{figure7}
\end{figure}

\begin{figure}[htb]
\centering
\includegraphics[width=.7\linewidth]{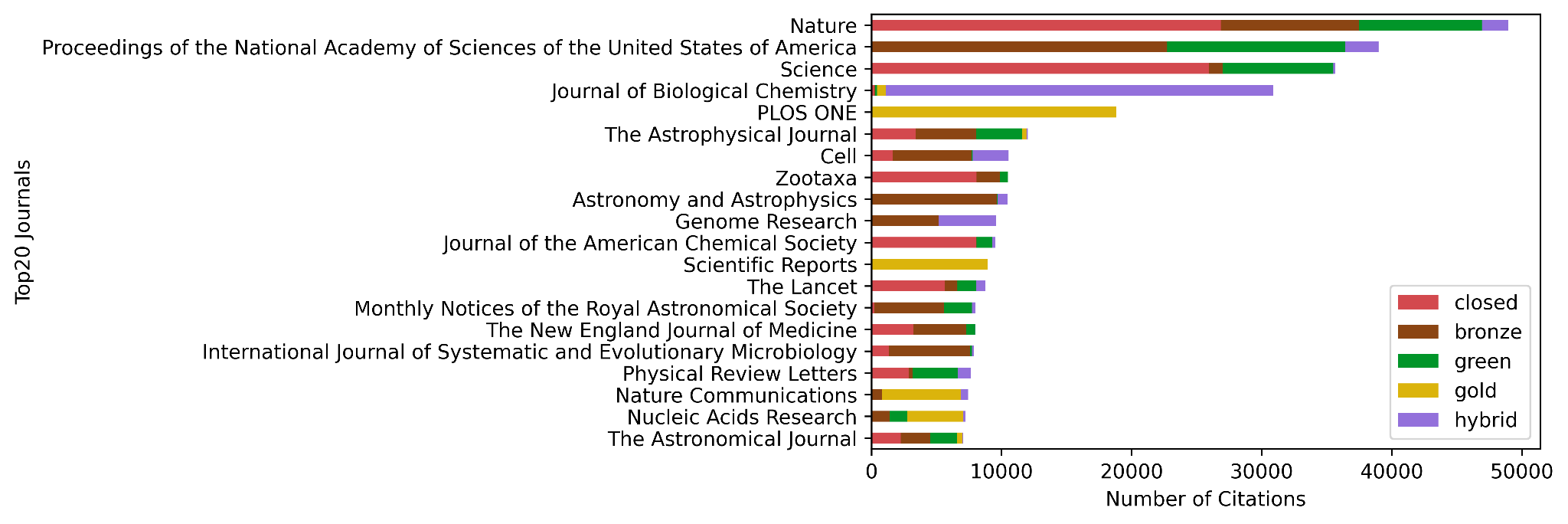}
\caption{Distribution of OA policies by top 20 journals.}
\label{figure8}
\end{figure}

\begin{figure}[H]
\centering
\includegraphics[width=.7\linewidth]{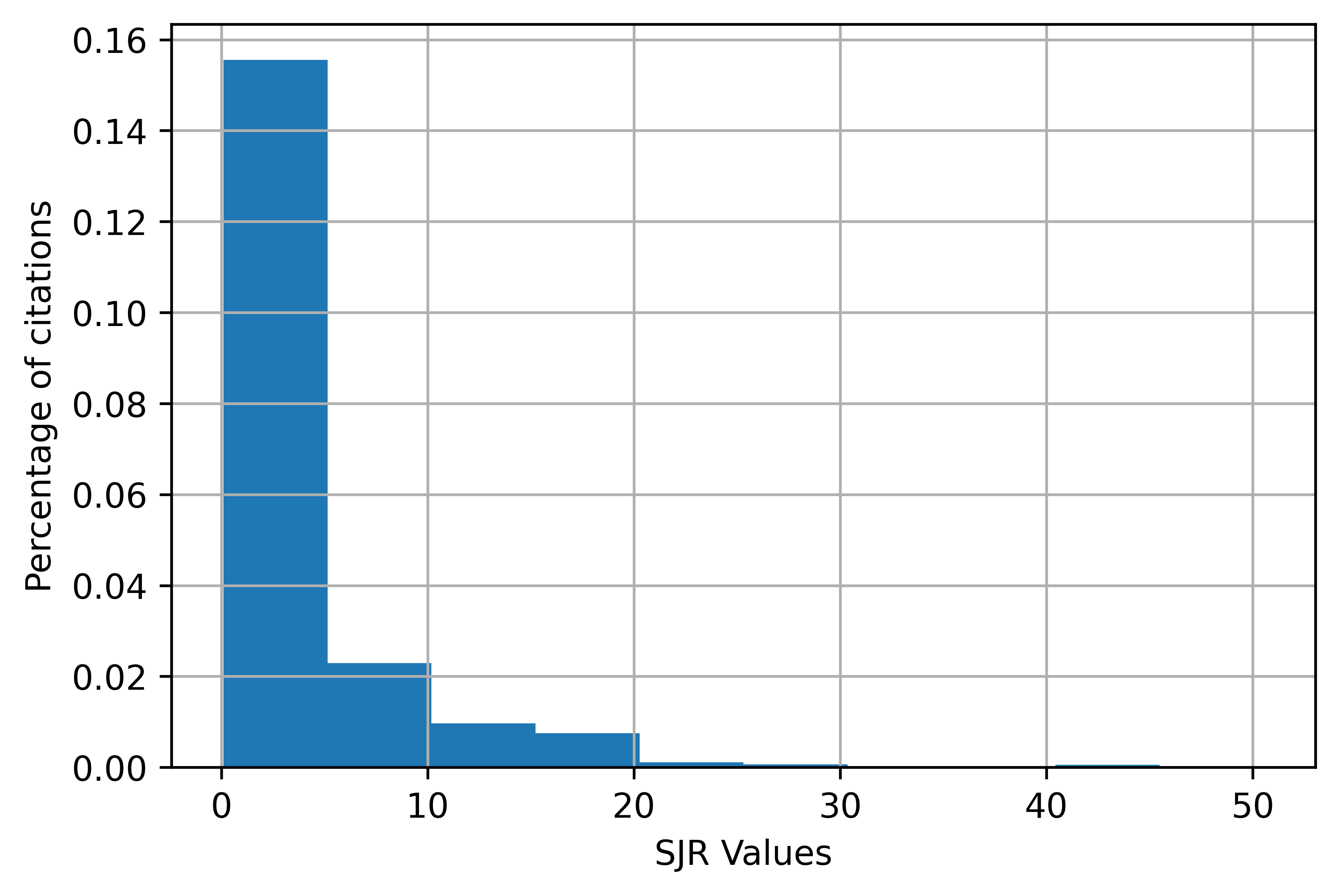}
\caption{Distribution of SJR among Wikipedia citations.}
\label{figure_SJR}
\end{figure}

\subsection{Tables}

The quality of our stratified samples is demonstrated through the descriptive statistics provided in Table \ref{tab:table15}, \ref{tab:table16}. Additionally, Table \ref{tab:table11} presents a count of articles by concepts within our dataset. The regression results for formulas 1 and 2 for the entire sample are displayed in Table \ref{tab:table12}, \ref{tab:table13}, \ref{tab:table5_all} and \ref{tab:table6_all}.

\begin{table}[ht]
\centering
\caption{Descriptive statistics for the articles cited in Wikipedia.}
\label{tab:table15}
\begin{tabular}{|l|c|c|c|c|c|c|}
\hline
      & cited\_by\_count & num\_references & article\_age & H index & is\_oa  & ln(SJR) \\ \hline
count & 261,230          & 261,230         & 261,230      & 261,230 & 261,230 & 261,230 \\ \hline
mean  & 146.133          & 37.797          & 252.171      & 249.348 & 0.504   & 0.632   \\ \hline
std   & 834.450          & 49.357          & 201.939      & 259.576 & 0.500   & 1.166   \\ \hline
min   & 0                & 0               & 12           & 0       & 0       & -2.303  \\ \hline
25\%  & 14               & 10              & 122          & 87      & 0       & -0.140  \\ \hline
50\%  & 45               & 28              & 208          & 164     & 1       & 0.554   \\ \hline
75\%  & 124              & 48              & 310          & 305     & 1       & 1.460   \\ \hline
max   & 304,415          & 1,976           & 1,487        & 1,331   & 1       & 3.922   \\ \hline
\end{tabular}
\end{table}

\begin{table}[htb]
\centering
\caption{Descriptive statistics for the articles not cited in Wikipedia. Average over all samples.}
\label{tab:table16}
\begin{tabular}{|l|c|c|c|c|c|c|}
\hline
      & cited\_by\_count & num\_references & article\_age & H index & is\_oa  & ln(SJR) \\ \hline
count & 261,230          & 261,230         & 261,230      & 261,230 & 261,230 & 261,230 \\ \hline
mean  & 59.008           & 28.316          & 252.137      & 249.348 & 0.497   & 0.632   \\ \hline
std   & 228.084          & 38.377          & 202.060      & 259.576 & 0.500   & 1.166   \\ \hline
min   & 0                & 0               & 12           & 0       & 0       & -2.303  \\ \hline
25\%  & 3                & 2               & 122          & 87      & 0       & -0.140  \\ \hline
50\%  & 18               & 21              & 208          & 164     & 0       & 0.554   \\ \hline
75\%  & 54               & 41              & 310.2        & 305     & 1       & 1.460   \\ \hline
max   & 44,132.4         & 2,891.2         & 1,487        & 1,331   & 1       & 3.922   \\ \hline
\end{tabular}
\end{table}

\begin{table}[htb]
\centering
\caption{Count of articles by concepts in the final combined dataset.}
\label{tab:table11}
\begin{tabular}{|c|c|c|c|c|c|}
\hline
Num. & Concept           & Counts  & Num. & Concept               & Counts \\ \hline
1    & Biology           & 144,307 & 11   & History               & 1,618  \\ \hline
2    & Medicine          & 48,286  & 12   & Art                   & 1,589  \\ \hline
3    & Chemistry         & 18,135  & 13   & Materials science     & 1,256  \\ \hline
4    & Physics           & 10,835  & 14   & Economics             & 1,005  \\ \hline
5    & Psychology        & 8,491   & 15   & Geography             & 911    \\ \hline
6    & Geology           & 8,429   & 16   & Business              & 480    \\ \hline
7    & Mathematics       & 5,620   & 17   & Sociology             & 466    \\ \hline
8    & Computer science  & 5,274   & 18   & Engineering           & 241    \\ \hline
9    & Philosophy        & 2,145   & 19   & Environmental science & 211    \\ \hline
10   & Political science & 1,931   &      &                       &        \\ \hline
\end{tabular}
\end{table}

\begin{table}[ht]
\centering
\caption{Coefficients for overall OA adoption. Average results across all 5 samples, model 1, $R^{2}=0.00032$.}
\label{tab:table12}
\begin{tabular}{|l|c|c|c|}
\hline
index                   & coef   & odds\_ratios & P\textgreater{}z \\ \hline
ln\_article\_age        & 0.064  & 1.066        & 0.000            \\ \hline
ln(SJR)                 & -0.002 & 0.998        & 0.482            \\ \hline
is\_oa                  & 0.583  & 1.791        & 0.000            \\ \hline
is\_oa:ln\_article\_age & -0.104 & 0.901        & 0.000            \\ \hline
\end{tabular}
\end{table}

\begin{table}[!t]
\centering
\caption{Coefficients for overall OA adoption. Average results across all 5 samples, model 2, $R^{2}=0.072$.}
\label{tab:table13}
\begin{tabular}{|l|c|c|c|}
\hline
index                   & coef   & odds\_ratios & P\textgreater{}z \\ \hline
ln1p\_times\_cited      & 0.442  & 1.556        & 0.000            \\ \hline
ln\_article\_age        & -0.068 & 0.934        & 0.000            \\ \hline
ln(SJR)                 & -0.245 & 0.782        & 0.000            \\ \hline
is\_oa                  & 0.494  & 1.639        & 0.000            \\ \hline
is\_oa:ln\_article\_age & -0.103 & 0.902        & 0.000            \\ \hline
\end{tabular}
\end{table}

\begin{table}[!htbp]
\centering
\caption{Coefficients for OA adoption by the policy. Average results across all 5 samples, model 1, $R^{2}=0.0013$.}
\label{tab:table5_all}
\begin{tabular}{|l|c|c|c|}
\hline
                        & coef   & odds\_ratios & P\textgreater{}z \\ \hline
Bronze                  & 0.194  & 1.215        & 0.002            \\ \hline
Gold                    & 0.672  & 1.959        & 0.000            \\ \hline
Green                   & 1.147  & 3.149        & 0.000            \\ \hline
Hybrid                  & 1.092  & 2.982        & 0.000            \\ \hline
ln\_article\_age        & 0.064  & 1.066        & 0.000            \\ \hline
Bronze:ln\_article\_age & -0.031 & 0.970        & 0.010            \\ \hline
Gold:ln\_article\_age   & -0.138 & 0.871        & 0.000            \\ \hline
Green:ln\_article\_age  & -0.189 & 0.828        & 0.000            \\ \hline
Hybrid:ln\_article\_age & -0.232 & 0.793        & 0.000            \\ \hline
ln(SJR)                 & -0.001 & 0.999        & 0.654            \\ \hline
\end{tabular}
\end{table}

\begin{table}[H]
\centering
\caption{Coefficients for OA adoption by the policy. Average results across all 5 samples, model 2, $R^{2}=0.073$.}
\label{tab:table6_all}
\begin{tabular}{|l|c|c|c|}
\hline
                        & coef   & odds\_ratios & P\textgreater{}z \\ \hline
Bronze                  & -0.158 & 0.855        & 0.015            \\ \hline
Gold                    & 1.071  & 2.920        & 0.000            \\ \hline
Green                   & 0.834  & 2.303        & 0.000            \\ \hline
Hybrid                  & 1.212  & 3.361        & 0.000            \\ \hline
ln1p\_times\_cited      & 0.447  & 1.564        & 0.000            \\ \hline
ln\_article\_age        & -0.071 & 0.932        & 0.000            \\ \hline
Bronze:ln\_article\_age & 0.038  & 1.039        & 0.002            \\ \hline
Gold:ln\_article\_age   & -0.241 & 0.786        & 0.000            \\ \hline
Green:ln\_article\_age  & -0.173 & 0.841        & 0.000            \\ \hline
Hybrid:ln\_article\_age & -0.284 & 0.753        & 0.000            \\ \hline
ln(SJR)                 & -0.244 & 0.784        & 0.000            \\ \hline
\end{tabular}
\end{table}

\subsection{Supplementary regression results}

This section provides an in-depth analysis of OA citation advantage for each OpenAlex concept. To achieve this, we developed 19 distinct regression models, each dedicated to analyzing the adoption of OA citation for a single concept. We use the second formulation for each model, with data pertaining solely to the corresponding concept being considered in each case.

To gain insight into the effect of OA adoption on each concept, we present the coefficients for the $\textit{is\_oa}$ variable in Table \ref{tab:table7} and the coefficients for the $\ln(1+\textit{times\_cited})$ variable in Table \ref{tab:table8}.

Table \ref{tab:table7} indicates that OA articles across most concepts exhibit a positive OA Wikipedia citation advantage, with five concepts showing statistically significant advantages. The top five concepts with the highest OA adoption advantage are Chemistry, Economics, Psychology, Business, and Physics, suggesting that STEM-related subjects attract more attention on Wikipedia.

Regarding $\ln(1+\textit{times\_cited})$ in each concept, citation counts demonstrate a significantly positive effect in nearly all concepts, although Environment Science and Engineering do not show significance. OA articles in several OpenAlex concepts, including Biology, Computer Science, Chemistry, Medicine, Psychology, Mathematics, Economics, Geology, Materials Science, and Physics, exhibit, on average, over a 30\% higher likelihood of being cited in Wikipedia compared to closed-access articles. Citation counts remain important factors in these concepts.

\begin{table}[!ht]
\centering
\caption{Coefficients for OA adoption by concept for all samples ($\textit{is\_oa}$).}
\label{tab:table7}
\begin{tabular}{|l|c|c|c|c|c|}
\hline
concept               & min OR & max OR & OR mean & Highest P-value & Mean R\textasciicircum{}2 \\ \hline
Biology               & 1.589  & 1.684  & 1.621   & 0.000           & 0.067                     \\ \hline
Computer science      & 1.314  & 1.737  & 1.471   & 0.264           & 0.052                     \\ \hline
Chemistry             & 3.828  & 4.068  & 3.918   & 0.000           & 0.060                     \\ \hline
Medicine              & 2.177  & 2.390  & 2.280   & 0.000           & 0.134                     \\ \hline
Psychology            & 2.524  & 3.554  & 3.150   & 0.001           & 0.090                     \\ \hline
Mathematics           & 1.536  & 1.748  & 1.672   & 0.227           & 0.073                     \\ \hline
Economics             & 2.277  & 3.635  & 3.250   & 0.276           & 0.104                     \\ \hline
Geology               & 1.347  & 1.531  & 1.428   & 0.136           & 0.053                     \\ \hline
Sociology             & 0.937  & 4.174  & 2.095   & 0.967           & 0.014                     \\ \hline
History               & 1.407  & 2.168  & 1.768   & 0.483           & 0.018                     \\ \hline
Geography             & 1.490  & 2.689  & 2.066   & 0.505           & 0.009                     \\ \hline
Philosophy            & 0.400  & 0.649  & 0.487   & 0.301           & 0.005                     \\ \hline
Materials science     & 1.511  & 3.379  & 2.253   & 0.497           & 0.098                     \\ \hline
Art                   & 0.783  & 1.175  & 0.961   & 0.902           & 0.008                     \\ \hline
Environmental science & 0.132  & 0.551  & 0.418   & 0.647           & 0.008                     \\ \hline
Physics               & 2.163  & 2.559  & 2.318   & 0.000           & 0.064                     \\ \hline
Engineering           & 0.863  & 1.965  & 1.313   & 0.911           & 0.006                     \\ \hline
Business              & 2.241  & 3.556  & 3.101   & 0.380           & 0.032                     \\ \hline
Political science     & 0.602  & 0.805  & 0.732   & 0.617           & 0.017                     \\ \hline
\end{tabular}
\end{table}

\begin{table}[ht]
\centering
\caption{Coefficients for OA adoption by concept for all samples ($\ln(1+\textit{times\_cited})$).}
\label{tab:table8}
\begin{tabular}{|l|c|c|c|c|c|}
\hline
concept               & min OR & max OR & OR mean & Highest P-value & Mean R\textasciicircum{}2 \\ \hline
Biology               & 1.594  & 1.602  & 1.599   & 0.000           & 0.067                     \\ \hline
Computer science      & 1.303  & 1.314  & 1.308   & 0.000           & 0.052                     \\ \hline
Chemistry             & 1.545  & 1.566  & 1.554   & 0.000           & 0.060                     \\ \hline
Medicine              & 1.820  & 1.835  & 1.824   & 0.000           & 0.134                     \\ \hline
Psychology            & 1.473  & 1.490  & 1.482   & 0.000           & 0.090                     \\ \hline
Mathematics           & 1.431  & 1.454  & 1.442   & 0.000           & 0.073                     \\ \hline
Economics             & 1.461  & 1.519  & 1.490   & 0.000           & 0.104                     \\ \hline
Geology               & 1.476  & 1.487  & 1.481   & 0.000           & 0.053                     \\ \hline
Sociology             & 1.113  & 1.211  & 1.153   & 0.003           & 0.014                     \\ \hline
History               & 1.252  & 1.297  & 1.274   & 0.000           & 0.018                     \\ \hline
Geography             & 1.124  & 1.149  & 1.137   & 0.000           & 0.009                     \\ \hline
Philosophy            & 1.109  & 1.126  & 1.116   & 0.000           & 0.005                     \\ \hline
Materials science     & 1.551  & 1.570  & 1.562   & 0.000           & 0.098                     \\ \hline
Art                   & 1.191  & 1.267  & 1.213   & 0.000           & 0.008                     \\ \hline
Environmental science & 1.011  & 1.149  & 1.069   & 0.824           & 0.008                     \\ \hline
Physics               & 1.409  & 1.418  & 1.413   & 0.000           & 0.064                     \\ \hline
Engineering           & 1.015  & 1.144  & 1.062   & 0.779           & 0.006                     \\ \hline
Business              & 1.205  & 1.240  & 1.221   & 0.000           & 0.032                     \\ \hline
Political science     & 1.203  & 1.240  & 1.227   & 0.000           & 0.017                     \\ \hline
\end{tabular}
\end{table}

\FloatBarrier
\bibliographystyle{unsrt}  
\bibliography{references}

\end{document}